\documentclass[useAMS,usenatbib]{mn2e}
\usepackage{color}
\usepackage{graphicx}
\usepackage{subfigure}
\usepackage{lscape}
\usepackage{amssymb,amsmath}

\title[Composite Circumstellar Dust Grains]{Composite Circumstellar Dust Grains}
\author[Gupta et al.]{Ranjan Gupta$^1$\thanks{E-mail:
rag@iucaa.in}, Dipak B. Vaidya$^2$ and Rajeshwari Dutta $^1$\\
1. IUCAA, Post Bag 4, Ganeshkhind, Pune-411007, India \\
2. Gujarat College, Ahmedabad-380006, India\\
}
\begin{document}

\date{Received on 2016 April 25; Accepted on 2016 July 13 }

\pagerange{\pageref{firstpage}--\pageref{lastpage}} \pubyear{2016}

\maketitle

\label{firstpage}

\begin{abstract}

We calculate the absorption efficiencies of composite silicate grains with inclusions of 
graphite and silicon carbide in the spectral range 5--25$\rm \mu m$.
We study the variation in absorption profiles with volume fractions of inclusions. 
In particular we study the variation in the wavelength of peak absorption at 10
and 18$\rm \mu m$. We also study the variation of the absorption of porous silicate grains.
We use the absorption efficiencies to calculate the infrared flux at
various dust temperatures and compare with the observed 
infrared emission flux from the circumstellar dust around some M-Type \& AGB stars obtained
from IRAS and a few stars from Spitzer satellite.
We interpret the observed data in terms of the circumstellar dust grain sizes;
shape; composition and dust temperature.

\end{abstract}


\section{Introduction}

In general, the composition of circumstellar dust around evolved stars are
studied from observations in the near and mid-infrared spectroscopy of absorption and
emission features. Emission from the characteristic 10 and 18$\rm \mu m$ features,
which arise from the
bending and stretching modes of silicate grains, were first identified in the spectra of oxygen rich giants
and super giants, \cite{woolf69}, \cite{woolf73}, \cite{bode88}.
 \cite{little} have analyzed about 450 IRAS-LRS spectra of M Mira variables to determine the
morphology of the emission features, found near 10$\rm \mu m$ and to correlate the shape of this
feature with period, mass loss rates and other parameters of the stars.
 \cite{simp} has analyzed the shape of the silicate dust features of 117 stars using
spherical dust shell models.
Evolved stars have distinctive IR spectra according to C/O abundance ratio e.g. 
\cite{aitken}, \cite{cohen}. It is to be noted, however, that among the stars
of the principal condensates in the
two environments, amorphous carbon and silicates, only those with silicate environments
have been detected directly. Amorphous carbon lacks strong IR resonances, although most of the
continuum emission from carbon stars is presumably attributed arising from amorphous
carbon, \cite{groen}. Silicon Carbide (SiC) emission at
11.3$\rm \mu m$ is the only spectral feature of dust commonly observed in normal C-type red giants.
However, as noted by \cite{lorenz}, SiC contributes only 10\% or less
of the dust in such objects. Although, the correlation between the C/O abundance ratio and the
form of the IR spectrum is not perfect, the dust features are sometimes used as diagnostic
of the C/O abundance ratios in stars. In some cases, silicate emission features are detected in stars
classified as C rich e.g. \cite{little86}, \cite{waters}.
Their analysis indicates that
the peak wavelength, strength and shape of the silicate features are very important to obtain
the exact composition, sizes and shapes of the silicate grains. Thus, in order to interpret
the observed silicate emission, we must compare the observed data with various
silicate based models. In this paper, we have systematically
analyzed the spectra of about 700 IRAS-LRS stars and compared them with composite dust grain models.
We have also analyzed four other M-type \& AGB stars observed by Spitzer satellite.

The grains flowing out of the stars are most likely to be
non-spherical and inhomogeneous, porous, fluffy and composites of many very small
particles due to grain-grain collisions, dust-gas interactions and various other processes.
Further, observations from space and balloon probes show that, in general, dust grains are porous,
fluffy and composites of many very small grains glued together, see \cite{brown}; 
\cite{kohler}; \cite{lasue} and \cite{levas}.
Since there is no exact theory to study the scattering properties of these
composite grains, various approximation methods are used for formulating models
of electromagnetic scattering by these grains, such as EMA (Effective Medium Approximation), DDA (Discrete
Dipole Approximation), etc. In EMA the optical properties (refractive index, dielectric
constant) of a small composite particle, comprising a mixture of two or more materials,
are approximated by a single averaged
optical constant and then Mie theory or T-Matrix is used to calculate absorption cross
sections for spherical/non-spherical particles. 
Basically, the inhomogeneous particle is replaced by a
homogeneous one with some average effective dielectric function.
The effects
related to the fluctuations of the dielectric function within the inhomogeneous
structures cannot be treated by this approach. For details on EMA, refer to 
\cite{bohren}.

On the other hand, DDA takes into account irregular shape
effects, surface roughness and internal structure of dust grains.
DDA is computationally more rigorous than EMA. (For a discussion and comparison of DDA
and EMA methods, including the limitations of EMA,
see \cite{bazel},
 \cite{perrin90a}, \cite{perrin90b}, \cite{ossen}
and \cite{wolff1994}. The DDA, which was first proposed by 
\cite{purcell}, represents a composite grain of arbitrary shape as a
finite array of dipole elements. Each dipole has an oscillating polarization in
response to both the incident radiation and the electric fields of the other
dipoles in the array, and the superposition of dipole polarizations leads to
extinction and scattering cross sections. For a detailed description of DDA, see
 \cite{draine1988}. 
In an earlier paper by \cite{vaidya2011}, the effects of inclusions and
porosities on the 10 and 18$\rm \mu m$ features were studied for the average
observed IRAS spectra.
In this paper, we use both DDA and EMA-T-Matrix
to study the absorption properties of the composite grains consisting of host silicate spheroidal
grains and inclusions of graphite or silicon carbide (SiC).
The effects of inclusions and
porosities, grain size and axial ratio (AR) on the absorption efficiencies of the
grains in the wavelength range 5--25 $\rm \mu m$ have been studied. In particular,
we have systematically studied the 10$\rm \mu m$ silicate feature as a function
of the volume fraction of the inclusions.
Using the absorption efficiencies of these composite grains for a power law
grain size distribution (\cite{mathis1977}),
the infrared fluxes for these grain models were calculated at various dust
temperatures (T=200--400K). The infrared flux curves obtained from the models
were then compared with the observed infrared emission curves of circumstellar
dust around 700 oxygen-rich M-type and AGB stars, obtained by IRAS
and Spitzer satellites. \cite{kessler2006}
have used opacities for distribution of hollow spheres (DHS) of silicate shells
 \cite{min05} and these models have been compared with circumstellar dust
around a few stars observed by Spitzer satellite. 
The DHS method averages the scattering and absorption/emission
cross sections of the set of hollow spheres. \cite{smolder}
have used silicates and gehlenite to study the 10 and 18$\rm \mu m$ peaks
in circumstellar dusts around S-type stars obtained by Spitzer satellite. 
Very recently, \cite{siber}
have used a mixture of amorphous carbon and silicate dust models to interpret interstellar
extinction, absorption, emission and polarization in the diffused interstellar medium.
\cite{mathis} and \cite{vosh2006} have used amorphous carbon with 
silicate in their composite grain model.
\cite{zubko} have reviewed various dust models including composite
silicate-graphite grain model and they have used EMA-T-Matrix method to calculate
extinction efficiencies composite grains. Using T-Matrix based method, \cite{iati}
have studied optical properties of composite interstellar grains. \cite{draine-li},
have used silicate-graphite-PAH model to study the infra-red emission from
interstellar dust in the post-Spitzer era. Using a radiative transfer model,
\cite{kirch} have studied the effect of dust porosity
on the appearance of proto-planetary disks.

A description of the composite grain models used is given in Section 2.
In Section 3, the results of the studies on the absorption efficiencies 
of the grain models are presented. Section 4 provides
results of the comparison of the model curves with the observed IR fluxes
obtained by IRAS and Spitzer satellites. The conclusions are presented in Section 5.

\section{Composite Grain Models}

The absorption efficiencies $\rm Q_{abs}$ of composite grains, made up of a host silicate
oblate spheroid and inclusions of graphite/silicon carbide and voids for porous silicate grains, for three axial ratios, in
the spectral region 5--25$\rm \mu m$, are computed using the DDA and EMA-T-Matrix
methods. The details on these composite grain models using DDA are provided in
\cite{vaidya2007}, \cite{vaidya2009} 
and \cite{vaidya2011}.

We have selected oblate (axial ratio $>$ 1) spheroid since,
it provides a good fit to the observed polarization across the 10$\rm \mu m$ feature
\cite{henning1993}, \cite{kim}.
Earlier, \cite{draine1984} have also used oblate silicate spheroids to 
model interstellar grains. \cite{gupta2005} have also shown that
the interstellar extinction curves obtained with oblate spheroidal grain models
fit best with the observed interstellar extinction.

{\it Discrete Dipole Approximation (DDA) based grain models:}

Composite spheroidal grain models containing N (number of dipoles) = 9640, 25896 and 14440 dipoles
(corresponding to axial ratios AR = 1.33, 1.5 and 2.0 respectively), each carved
out from $32 \times 24 \times 24, 48\times 32\times 32$ and $48\times 24\times 24$
dipole sites respectively are used.
Sites outside the spheroid are set to be vacuum and sites inside are assigned to
be the host material. The size of the inclusion is given in terms of the number of
dipoles 'n' across the diameter of an inclusion, e.g. n = 152 dipoles for composite
grain model with N = 9640, i.e. AR = 1.33 and volume fraction of f=0.1. 
Please see Table 1 in \cite{vaidya2011} and also
 \cite{chylek2000}. 
The DDA code generates a three dimensional matrix, specifying
the material type at each dipole site. 
There are two validity criteria for DDA (see e.g. \cite{wolff1994});
viz. (i) $\rm |m|kd \leq 1$, where m is the complex refractive index
of the material; k=$\rm \pi/\lambda$ is the wavenumber and
 d is the lattice dispersion spacing and
(ii) d should be small enough (N should be sufficiently large) to
 describe the shape of the particle satisfactorily.
For all the composite grain models, viz. N=9640; 14440 \& 25896; and for all
grain sizes 0.005-0.250$\mu $ in the wavelength range 5--25$\rm \mu m$ considered in the
present study, we have checked that the DDA validity criteria are satisfied.
Details on the computer code and the
corresponding modification to the DDSCAT 6.1 code 
\cite{draineflatau2003} are given in \cite{vaidya2001} and \cite{gupta2005}.

{\it Effective Medium Approximation (EMA-T-Matrix) based models:}

We have also obtained absorption efficiencies for composite grains
using the combination of EMA-T-Matrix calculations. In EMA, a mixing rule
(e.g. Maxwell-Garnet or Bruggman see: 
 \cite{bohren}) is used to obtain the average refractive
index (by using the optical constants of two materials) of the composite grain.
We have used Maxwell-Garnet mixing rule. The absorption efficiencies of the composite grains
are calculated using the T-Matrix code given by \cite{mish2002}.
Earlier, \cite{gupta2005} and \cite{vaidya2009} have used EMA-T-Matrix based method
to study scattering properties of composite grains.
For a discussion on various EMA mixing rules see \cite{bohren}.

{\it Grain size and size distribution:}

We have used the power-law MRN, \cite{mathis1977}, dust grain size distribution:

n(a) $\rm \propto a^{-q} (a_{min} < a < a_{max})$,

where a is the effective radius. This law states that the
number of dust grains per unit volume having radius between a and a + da, is proportional to
$a^{-q}$. An acceptable fit to the observed
data is obtained with values of $\rm a_{min} = 0.005 \mu, a_{max} = 0.250 \mu$ and q = 3.5.
If the semi-major and semi-minor axes are denoted by x/2 and y/2 respectively,
then $\rm a^3=(x/2)(y/2)^2$, where 'a' is the radius of the equivalent sphere,
whose volume is the same as that of the spheroid.

{\it Types of Inclusions:}

For the composite grain models, we have used the host silicate spheroid with graphite 
or silicon carbide (SiC) inclusions.
Graphite is selected as inclusions, since it explains the observed bump at 2175\AA~ in the
UV region of the extinction curve \cite{draine1988}.
\cite{odon} and \cite{henning1993} have also used graphite
inclusions in the composite grain models.
\cite{min07}, \cite{min08} have used SiC as inclusions to study the 10$\rm \mu m$
silicate feature.

We have also used porous silicate grain models, as the
circumstellar grains are likely to be porous and fluffy, 
 \cite{kohler} and \cite{vosh2005}.

The complex refractive indices for silicates and graphite were obtained
from \cite{draine2003}. The optical constants for SiC were obtained from 
 \cite{pegourie1988}.
The volume fractions of the inclusions are denoted as f = 0.04, 0.06, 0.08, 0.1, 0.2 \& 0.30 etc.

{\it Orientation:}

For randomly oriented spheroidal grains, the scattering properties of the
composite grains have to be averaged over all of the possible orientations. In the present study,
we have used three values for each of the orientation parameters 
and averaging is done over 27 different orientations for all DDA 
computations (see \cite{wolff1994} \& \cite{wolff1998}).

For an illustration of a composite grain model with N = 14440 (total 14440 dipoles
with 11 inclusions of 224 dipoles per inclusion); see Fig.1a and Fig1b.

\begin{figure}
  \centering
  \subfigure{\includegraphics[width=0.4\textwidth]{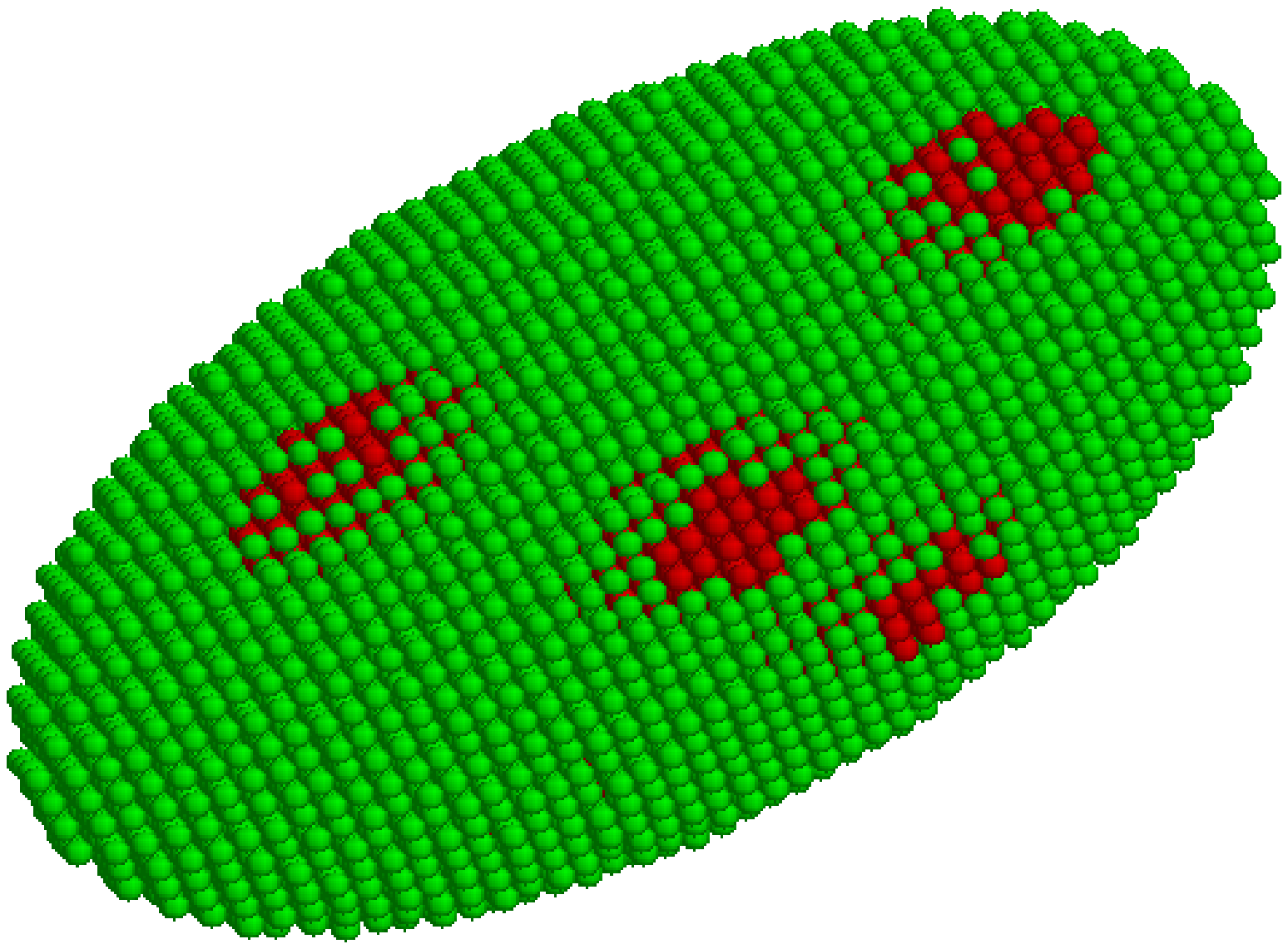}\label{fig:f1a}}
  \subfigure{\includegraphics[width=0.4\textwidth]{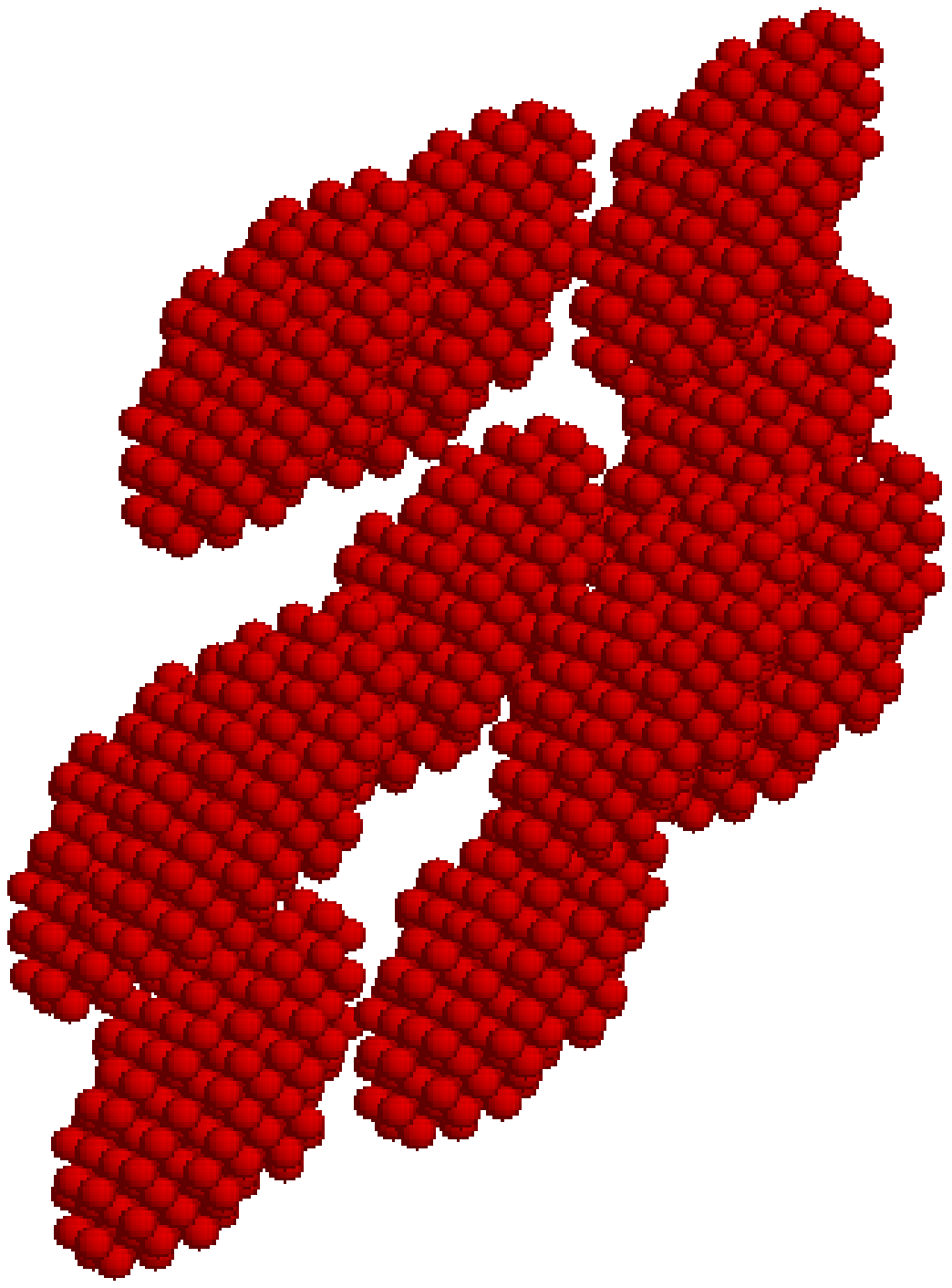}\label{fig:f1b}}
\caption{Figure 1a(top): A non-spherical composite dust grain consisting of host (in green) and
inclusion (in red) with a total of N=14440 dipoles where the 11 inclusions are 
embedded in the host spheroid are shown such that only the ones placed at the outer 
periphery are seen in this panel. 
Figure 1b(bottom): The 11 inclusions are shown separately in this panel (somewhat enlarged than Fig. 1a).}
\end{figure}

\section {Model Absorption Curves and Absorption Efficiency of Composite Grains}

We have calculated the absorption efficiencies for the composite grain models
with spheroidal silicates as host and graphite/or silicon carbide inclusions.
The composite grain models with
graphite inclusions have been used earlier by \cite{ossen}, \cite{odon},
\cite{henning1993}, \cite{vaidya2011} and 
with SiC by \cite{kessler2006}, \cite{min08} 
and \cite{siber}.
We have also calculated the absorption efficiency for the porous silicate grains.
Porosity 'P' is defined as $\rm P=1-V_{solid}/V_{total}$; where
$\rm V_{solid}$ is the number of dipoles of the material and $\rm V_{total}$
is the total number of dipoles in the spheroid; accordingly the porosity
varies from 1--30\% for the spheroidal grains (\cite{green} and
\cite{vaidya1997}).
Earlier, porous silicate grains have also been used by \cite{li} for modeling
the dust around AGN stars. \cite{vosh2013} have studied the effect of porosities
on the 10$\rm \mu m$ silicate feature. Using \cite{kohler}, have calculated
radiation pressure cross-sections on porous (porosities 30--90\%) astronomical spherical silicate
grains in the circumstellar dust shells around $\beta$-Pictoris, Vega \& Formalhaut stars.
\cite{kohler} have also used the combination of EMA and Mie theory and 
calculated the radiation pressure cross-sections of porous silicate grains and compared with
the DDA results. They have also noted that the DDA method is time consuming.

The variation in absorption efficiencies $\rm Q_{abs}$ of the composite
grains with different
volume fractions (0.04, 0.10,...0.3) of the inclusions of Silicon Carbide (SiC) 
for the three different axial ratios 1.33, 1.5 and 2.0, in the wavelength range 5--25$\rm \mu m$
for the grain size a = 0.10$\rm \mu$ is shown in Figs. 2(a, b \& c).
It is seen that for the composite grain models with inclusions of SiC,
the strength of both the 10 and 18$\rm \mu m$ absorption peaks decreases with 
increase in the volume fractions. It is to be noted that for higher volume fractions
0.3 of SiC inclusions, the the 10$\rm \mu m$ feature is modified considerably.

\begin{figure}
\includegraphics[width=84mm]{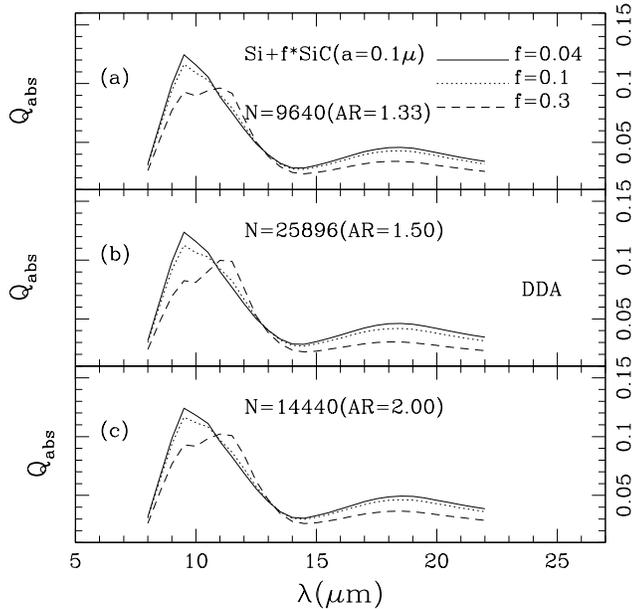}
\caption{Variation in $\rm Q_{abs}$ with SiC inclusions 
for three axial ratios and grain size 0.10$\rm \mu$ over
wavelength range 5–-25$\rm \mu m$ using DDA calculations.}
\end{figure}

We have also computed absorption efficiencies of the composite grains
using EMA-T-Matrix based calculations. In Figs. 3(a, b \& c), we show the absorption profiles of 
composite grains with three SiC inclusions. 
It is seen that the absorption efficiencies increase with volume fraction of SiC and there 
is no appreciable shift in the wavelength of peak absorption at 10$\rm \mu m$ feature.
In our earlier study \cite{vaidya2011}, using DDA, we have shown the variation of
absorption efficiencies of composite grains with inclusions of graphite and porous
silicate grains.

\begin{figure}
\includegraphics[width=84mm]{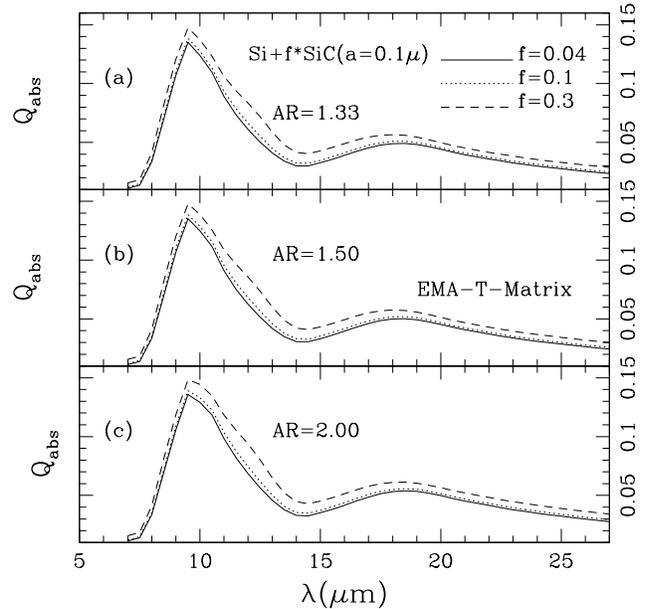}
\caption{Variation in $\rm Q_{abs}$ with SiC inclusions 
for different axial ratios (AR) and grain size 0.10$\rm \mu$ over wavelength range 5--25$\rm \mu m$,
using EMA-T-Matrix calculations}
\end{figure}

In Fig.4, we compare the absorption efficiencies of the composite grains with inclusions of
graphite and porous grains 
obtained using DDA and EMA-T-Matrix methods. 
It is seen in Fig 4 (a) \& (c) the absorption efficiencies decrease as the volume fractions
of graphite increases for both DDA \& EMA-T-Matrix. The absorption efficiencies for porous
silicate grains as seen in Fig. 4 (b) for DDA decreases with increasing 
inclusion fraction whereas for EMA-T-Matrix in 4 (d) there is no appreciable change.
Figure 5 shows how the ratio between $\rm Q_{abs}$ calculated from DDA and EMA-T-Matrix
deviate and it clearly shows that for higher volume fractions, the ratio deviations are more
prominent around the 15$\rm \mu m$ region.

The absorption efficiencies obtained using EMA-T-Matrix calculations and DDA do not agree because
the EMA method does not take into account the inhomogeneities within the grains
(internal structure, surface roughness etc see \cite{wolff1998} and the material
interfaces and shapes are smeared out into the homogeneous 'average mixture'; thus
the refractive index using EMA is an average one and the resulting absorption efficiency is 
not reliable, \cite{saija2001}. However, EMA-T-Matrix method 
is still quite useful since the application of DDA poses a computational challenge for
large size parameter $\rm X=2 \pi a/\lambda > 20$ and large complex refractive index m.
Further, EMA allows to examine applicability of several mixing rules see e.g.
\cite{wolff1998}; \cite{chylek2000}; \cite{saija2001};
\cite{vosh2005} and \cite{vosh2006}.

\begin{figure}
\includegraphics[width=84mm]{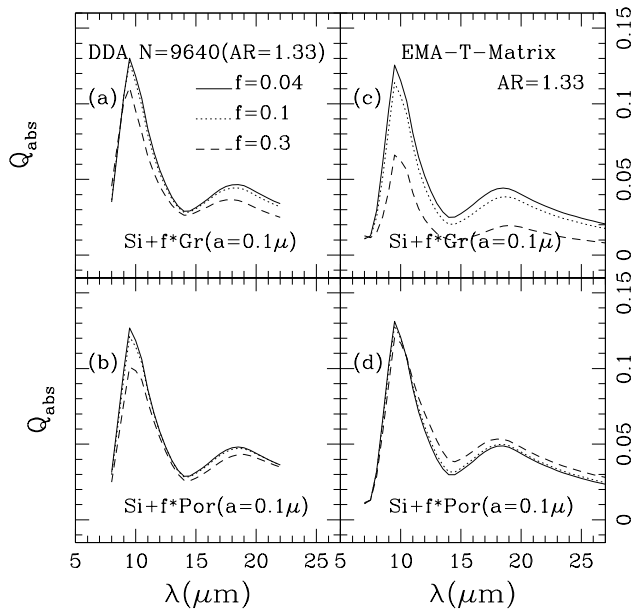}
\caption{Absorption efficiencies of composite grains with graphite inclusions (a) \& (c) \& 
porous silicates (b) \& (d), using DDA \& EMA-T-Matrix methods}
\end{figure}

\begin{figure}
\includegraphics[width=84mm]{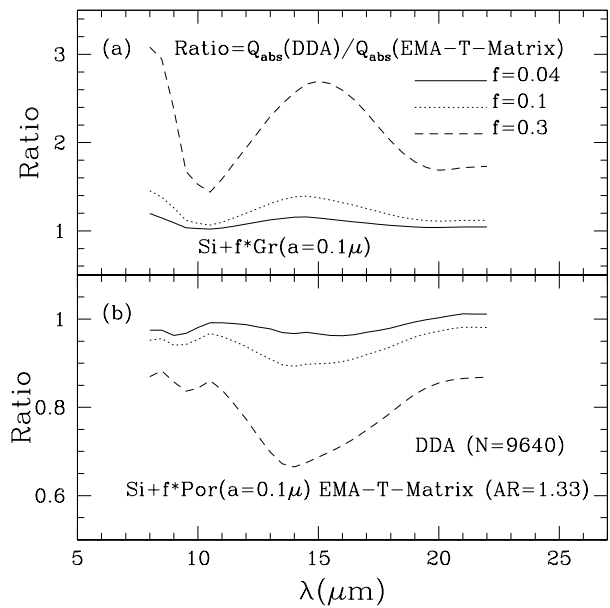}
\caption{Ratio of Absorption efficiencies of composite grains with graphite inclusions (panel (a)) 
\& porous silicates (panel (b)), using DDA \& EMA-T-Matrix methods}
\end{figure}

\section {IR Emission from Dust and Comparison with Observed Data}

The stars which have evolved from the main sequence and have entered the giant phase of their evolution
are a major source of dust grains in the galactic interstellar medium. Such stars have 
oxygen-rich material
(e.g. olivine) over abundant relative to carbon, and therefore, produce silicates and show
strong feature at 10$\rm \mu m$.
The 10$\rm \mu m$ feature is identified with the Si--O bond, from stretching mode.
These materials (amorphous silicates, e.g. olivines and pyroxenes) also show a weaker feature at 
18$\rm \mu m$ resulting from O--Si--O bending mode, \cite{little} and \cite{whitt}.
These features can be present in either
emission or absorption spectra depending on the optical depth of the circumstellar
shell. 

The Infrared Astronomical Satellite (IRAS) had an instrument called the Low
Resolution Spectrometer (LRS), which measured spectra of the brighter ($>$ 10Jy)
point sources (about 50,000), between 7.7 and 22.6$\rm \mu m$, with a resolution
varying from 20 to 60. A total of 5425 objects with better quality spectra were
included in the Atlas of Low-Resolution IRAS Spectra, \cite{olnon}.
Two thousand bright sources from the Atlas were classified into 17 different
classes based on their spectral morphology using Artificial Neural Network (ANN)
scheme \cite{gupta2004}. Objects belonging to Class 6 of this classification,
which are O--rich AGB stars with strong silicate emission feature at 10$\rm \mu m$, are
considered for this paper. This class contains the largest number of objects
(732) amongst all the classes.

Using the absorption efficiencies $\rm Q_{abs}$ of the composite grains, the infrared flux,
$\rm F_{\lambda}$ at various dust temperatures is calculated using the relation,
$\rm F_{\lambda} = Q_{abs}  B_{\lambda}(T)$
at dust temperature T in K, and $\rm B_{\lambda}$ is the Planck function.
This relation is valid only if
the silicate emission region is optically thin, \cite{simp}, \cite{ossen92}, \& \cite{li}.
It is also assumed that the dust is isothermal and the grains behave as
perfect black bodies.
As mentioned earlier in Section 2, we have used a MRN (power law) grain size distribution
with $\rm a_{min}=0.005\mu$ and $\rm a_{max}=0.250\mu$, \cite{mathis1977}
for all the composite grain models to compare the observed IR fluxes.
The composite grain model with larger grain size distribution in the range a=0.1 to 1.0$\rm \mu$
did not match the observed curves satisfactorily. As noted by \cite{sylv},
 \cite{teles}, \cite{back}, \cite{li1998}, \cite{kriv} and \cite{car}, small
grain sizes are expected from the IR data and the profile of the silicate emission at 10$\rm \mu m$.
In Figure 6, we show the variation of the infrared flux for a temperature range of
200--400K for composite grains with 10\% SiC inclusions, and size integration of grain sizes from 
0.005 to 0.250 $\rm \mu$ by both DDA amd EMA-T-Matrix methods. 
It is seen that the flux decreases as the dust temperature increases.

\begin{figure}
\includegraphics[width=84mm]{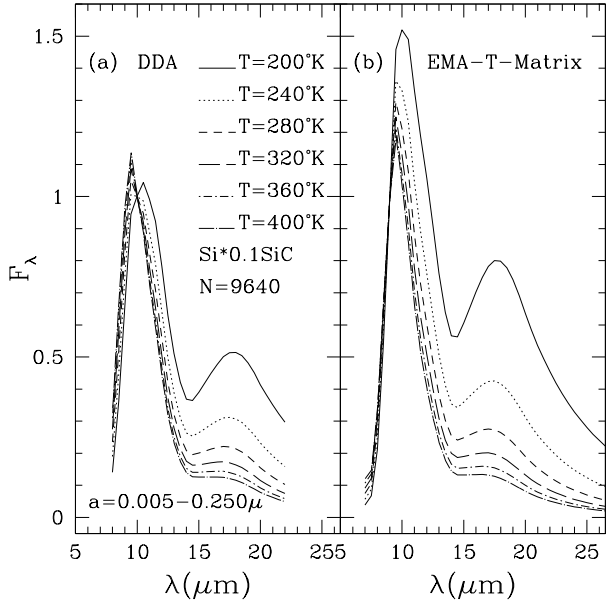}
\caption{Infrared flux at various temperatures for composite grains with SiC inclusions.}
\end{figure}

\begin{figure}
\includegraphics[width=84mm]{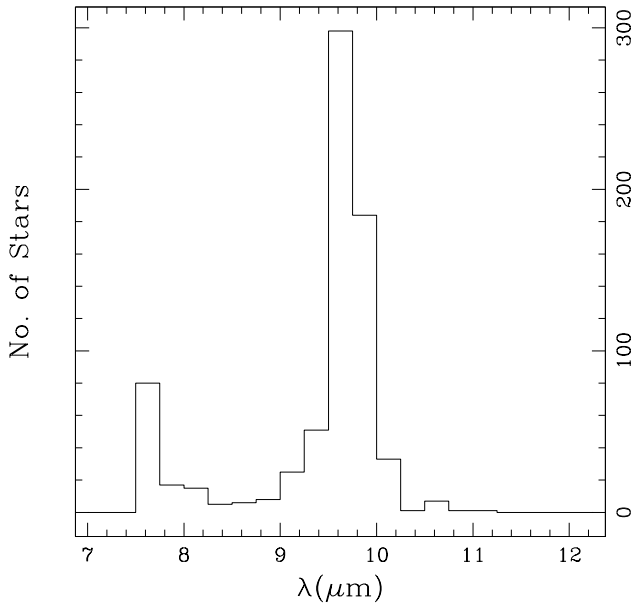}
\caption{Histogram showing the statistics of the 10$\rm \mu m$ feature's variation amongst
the 700 IRAS objects.}
\end{figure}

The observed IRAS-LRS curves obtained for circumstellar dust around 700
oxygen rich M-type stars, are compared with the calculated infrared fluxes, $\rm F_{\lambda}$,
for the composite grain models. $\chi^{2}$ minimization method is used to fit the observed
data with the calculated infrared fluxes for the composite grain models
models. For details on the $\chi^{2}$ method see \cite{vaidya2007} \& \cite{vaidya2011}.
An error bar of 10\% on the data is used, \cite{olnon}.
In Fig. 7 histogram, we show the number of stars with the absorption band at 10$\rm \mu m$.
It is seen that the maximum number of stars have peak wavelength range between 9.4 \& 9.8 $\rm \mu m$.

The best fit dust temperature is computed for each star along with the errors in
temperature over a confidence level of 90\%. 

Some of the best fit infrared emission flux curves, for the composite grain
models are shown in Fig. 8 with graphite; porous and SiC inclusions \& porous silicate grains.
The smooth curves represent the model data, while the points with error bars
represent the observed data.
The axial ratio and the dust temperature for the best fit grain model
are also shown in the plots. 

\begin{figure}
\includegraphics[width=84mm]{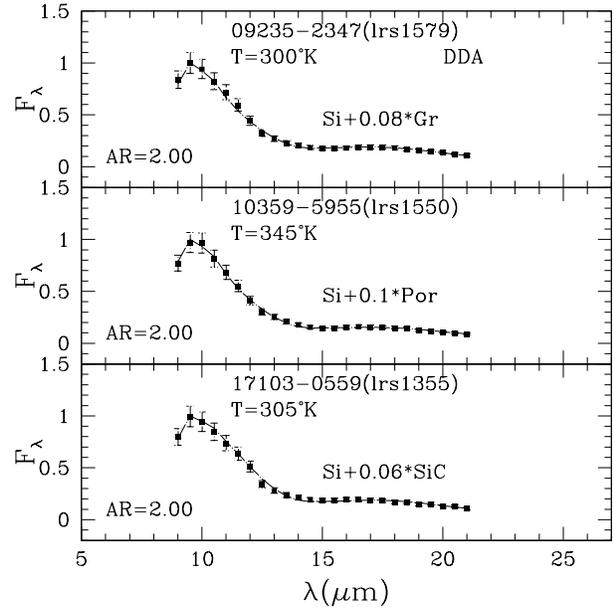}
\caption{Composite grain models (DDA) with various inclusions fits to IRAS-LRS data.
The number at top of each panel is the IRAS identifier.}
\end{figure}

We also compare the observed IRAS-LRS flux data with
best fit models using EMA-T-Matrix calculations in Fig. 9.

\begin{figure}
\includegraphics[width=84mm]{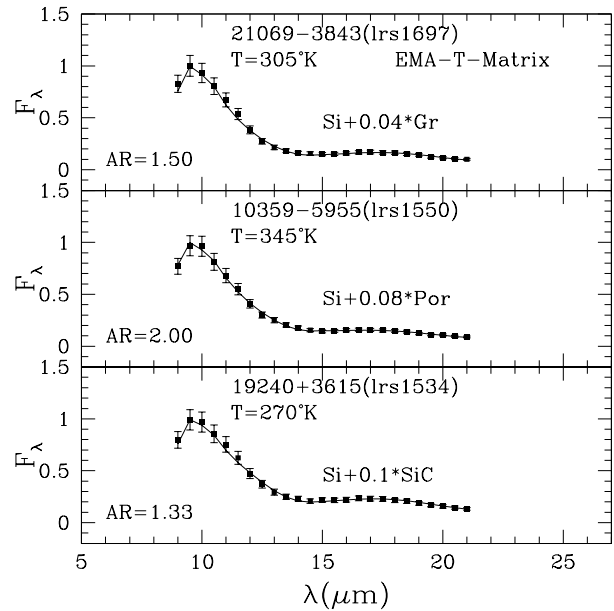}
\caption{Composite grain models (EMA-T-Matrix) with various inclusions fits to IRAS-LRS data.
The number at top of each panel is the IRAS identifier.}
\end{figure}

Table 1 shows the detailed parameters for composite grain and porous silicate grain
models for some of the best fits of these models to a set of 10 IRAS stars, each with their
IRAS catalogue No.; corresponding LRS No.; the
best fit minimum $\chi^{2}$ value; estimated temperature; model evaluated flux ratio 
(R=Flux(18$\rm \mu$)/Flux(10$\rm \mu$));
observed flux ratio (R) and the corresponding best fit DDA model. 
Table 2 shows the best fit results with EMA-T-Matrix model.

\begin{table*}
\caption{List of best $\chi^{2}$ fit DDA based models to the IRAS objects -- a list of best
selected 10 stars. The best fit model notation e.g. 2.00si0.08gr denotes that
the axial ratio (AR) is 2 and si0.08gr denotes 8\% volume inclusion of graphite 
in the host silicate grain.}
\centering
\begin{tabular}{|l|c|c|c|c|c|c|}
\hline
IRAS No. & LRS No. & Min.$\chi^{2}$ & T(K) & Model R & Obs.R & BestFitModel \\
\hline
09235-2347 & lrs1579  & 0.1515 & 300.0 & 0.180 & 0.180 & 2.00si0.08gr\\
17119+0859 & lrs1915  & 0.1710 & 285.0 & 0.202 & 0.203 & 2.00si0.08gr\\
10359-5955 & lrs1550  & 0.1863 & 345.0 & 0.146 & 0.149 & 2.00si0.1por\\
20484-7202 & lrs1670  & 0.1905 & 295.0 & 0.186 & 0.176 & 2.00si0.1gr\\
18595-3947 & lrs1975  & 0.1963 & 290.0 & 0.182 & 0.173 & 2.00si0.3gr\\
19240+3615 & lrs1534  & 0.2002 & 275.0 & 0.218 & 0.213 & 2.00si0.1gr\\
19244+1115 & lrs1992  & 0.2057 & 220.0 & 0.390 & 0.379 & 2.00si0.2gr\\
11525-5057 & lrs1482  & 0.2189 & 295.0 & 0.187 & 0.178 & 2.00si0.08gr\\
16340-4634 & lrs1946  & 0.2208 & 265.0 & 0.232 & 0.234 & 2.00si0.2gr\\
06297+4045 & lrs1617  & 0.2225 & 260.0 & 0.244 & 0.244 & 2.00si0.2gr\\
\hline
\end{tabular}
\end{table*}

\begin{table*}
\caption{List of best $\chi^{2}$ fit EMA-T-Matrix based models to the IRAS objects -- a list of best
selected 10 stars. The best fit model notation e.g. 2.00si0.08por denotes that
the axial ratio (AR) is 2 and si0.08por denotes 8\% volume inclusion of vacuum
in the host silicate grain making it a porous silicate grain.}
\centering
\begin{tabular}{|l|c|c|c|c|c|c|}
\hline
IRAS No. & LRS No. & Min.$\chi^{2}$ & T(K) & Model R & Obs.R & BestFitModel \\
\hline
10359-5955 & lrs1550 & 0.1500 & 345.0 & 0.146 & 0.149 & 2.00si0.08por \\
09235-2347 & lrs1579 & 0.1887 & 310.0 & 0.180 & 0.180 & 2.00si0.08por \\
17119+0859 & lrs1915 & 0.2043 & 285.0 & 0.205 & 0.203 & 1.50si0.08por \\
02351-2711 & lrs1884 & 0.2160 & 330.0 & 0.159 & 0.156 & 2.00si0.08por \\
19240+3615 & lrs1534 & 0.2405 & 270.0 & 0.222 & 0.213 & 1.33si0.1sic \\
11525-5057 & lrs1482 & 0.2429 & 295.0 & 0.190 & 0.178 & 1.50si0.08por \\
17484-0800 & lrs1712 & 0.2495 & 315.0 & 0.174 & 0.166 & 2.00si0.08por \\
15255+1944 & lrs1764 & 0.2528 & 290.0 & 0.194 & 0.194 & 1.33si0.08por \\
08124-4133 & lrs1734 & 0.2544 & 295.0 & 0.191 & 0.188 & 1.50si0.3sic \\
21069-3843 & lrs1697 & 0.2556 & 305.0 & 0.162 & 0.164 & 1.50si0.04gr \\
\hline
\end{tabular}
\end{table*}

The results in Figs. 8 \& 9, and Tables 1 \& 2, show that the composite grain models with 
graphite inclusions fit the observed fluxes reasonably well,
whereas the porous silicate models and the composite grain models with inclusions of SiC
do not fit the observed curves satisfactorily, 
as indicated by the $\chi^{2}$ values.

Further, these results indicate that the temperature range of 250--350K derived from the
grain models give satisfactory fit to the observed data. Stars giving good fits with the
composite grain models with graphite \& SiC inclusions have temperatures between 270 and 300K, while
those with the porous silicate grain models give the best fit dust temperature between
300 and 370K. The estimated temperature errors are within 10K.

We have also used the data from the Infrared Spectrograph IRS (\cite{houck});
on board the Spitzer Space Telescope, \cite{werner} and extracted the spectrum of
four M-type and AGB stars using the Cornell Atlas of Spitzer/IRS Sources (CASSIS)
extraction procedure (\cite{lebout}). Figure 10 shows model fits with observed Spitzer data in the
spectral range of 8-13$\rm \mu m$ for M-type stars and 
one AGB star (see \cite{mant}).
It is seen that the composite grain models or porous silicate grain models do not fit
the observed Spitzer fluxes satisfactorily.
The derived temperature range in these objects is around 220--250K.
The stars in panels (a) and (b) fit to the composite grains with graphite inclusions and porous silicates
respectively whereas the stars in panels (c) and (d) show the best fit to the composite grain
models with SiC inclusions and
seem to have been affected by the SiC feature at 11.3$\rm \mu m$.
\cite{smolder} have used silicate grains with gehlenite 
grain models to fit the observed spectra
of a few stars obtained by Spitzer, and have also noted that grain models with a component
of SiC are not suitable to fit the M-type stars.
We note here that we need to analyze larger sample of M-type and AGB stars observed by
Spitzer satellite. 

\begin{figure}
\includegraphics[width=84mm]{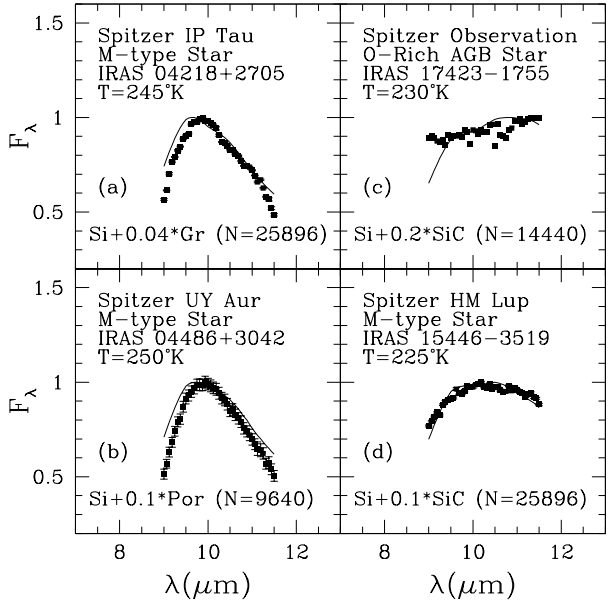}
\caption{Best fits to a set of Spitzer observed spectra. The solid points with error bars
are the Spitzer data and continuous line is the best fit model.}
\end{figure}

\cite{min08} have used grain models with mixture of SiC and Si 
to interpret the observed 10$\rm \mu m$ absorption band
and they have noted that the shape of SiC resonance is very sensitive to the 
shape of the SiC grain in the region of 10$\rm \mu m$ band. 
In Figs. 11 (a), (b) \& (c) we show the effect of grain shape (axial ratio AR) on the
silicate and SiC features at 10 and 11.3$\rm \mu m$ respectively.
It is seen that the 10.0$\rm \mu m$ silicate feature is not affected by the grain
shape, whereas, 11.3$\rm \mu m$ SiC feature shows variation with the grain shape (see
Fig. 11 b \& c).

\begin{figure}
\includegraphics[width=84mm]{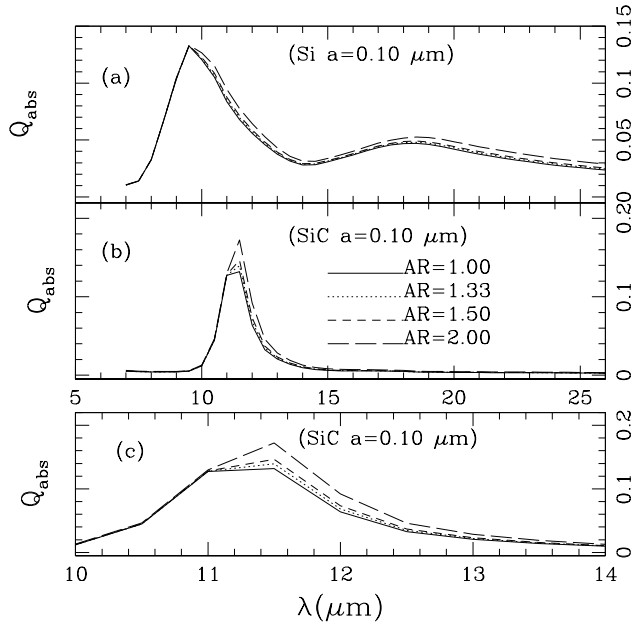}
\caption{The panels (a) and (b) show the absorption efficiencies $\rm Q_{abs}$
at various axial ratios (AR) for Silicates (Si) and Silicon Carbide (SiC).
The x-axis for both these panels is wavelength $\lambda$ in $\rm (\mu m)$.
The panel (c) is same as (b) but with the wavelength axis expanded to highlight
the variation in the shape of the peak absorption feature of SiC at 11.3$\rm \mu m$ with 
the grain shape (i.e. AR).}
\end{figure}

{\it Flux Ratio R in the Silicate Emission:}

We have also calculated the flux ratio
R=Flux(18$\rm \mu$)/Flux(10$\rm \mu$) for all the composite grain models.

In our earlier study \cite{vaidya2011},
we had shown that, in general, the flux ratio R decreases
with the dust temperature for the composite grains with graphite inclusions and
porous silicate grain models, it varies between $\sim$0.6 at 200K
and 0.2 at 300K. We had also shown that for the composite grain models with graphite
inclusions, the ratio R decreases with the volume fraction of the inclusions, whereas it increases
with the porosities. \cite{henning1993} did not find any variation either
with graphite inclusions or with the porosities. 
In Tables 1, 2 \& 3, we show the flux ratio for the best fit models.
In this paper, we also show R for the composite grains with SiC inclusions as depicted in Table 3.
It is seen that the R decreases with the dust temperature which is consistent with the earlier
results on composite grains with graphite inclusions and porous silicate grains,
\cite{vaidya2011}. We also note that R obtained by EMA-T-Matrix is higher 
than that obtained by DDA.

\begin{table*}
\caption{Variation of the flux ratio R=Flux(18$\rm \mu$)/Flux(10$\rm \mu$) for various
temperatures for SiC inclusions with DDA and EMA-T-Matrix models.}
\centering
\begin{tabular}{|l|c|c|}
\hline
T(K) & DDA & EMA-T-Matrix\\
\hline
200 & 0.501 & 0.702\\
240 & 0.310 & 0.397\\
280 & 0.221 & 0.266\\
320 & 0.173 & 0.198\\
360 & 0.144 & 0.156\\
400 & 0.124 & 0.133\\
\hline
\end{tabular}
\end{table*}

\section {Conclusions}

In order to interpret dust properties (viz. sizes, shapes, composition and temperature)
in circumstellar dust, we have
systematically analyzed the IR spectra of about 700 IRAS-LRS oxygen rich M type stars using the
composite grain models with graphite and silicon carbide (SiC) inclusions
and porous silicate grain models. The 10$\rm \mu m$ band has been studied in
more details and is compared with the observed data for the stars with the absorption band between
9.5 and 10.0$\rm \mu m$. It should be noted here that we have used both DDA \& EMA-T-Matrix
methods to compute the absorption efficiency of composite grain models and porous 
silicate grain models, whereas most of others have used EMA
and T-Matrix methods or methods using Distribution of Hollow Spheres (DHS),
\cite{min08} and \cite{kessler2006} or layered spheres method \cite{mathis}.

The main conclusions are presented below:

(1) The absorption/emission properties of the silicate grains are modified considerably
with the inclusions
of carbonaceous materials (e.g. graphite \& SiC) and porosity. Our results on the
composite grain models clearly show the variation in absorption efficiency $\rm Q_{abs}$ with the
variation of volume fraction of inclusions (graphite \& SiC) and porosity. 
The results on the composite
grain models with inclusions of SiC show a longward shift in the 10$\rm \mu m$ absorption feature
with the volume fraction of the inclusions whereas the composite grain models with graphite
inclusions show a shortward shift in the 10$\rm \mu m$ feature.
The 18$\rm \mu m$ feature, however, remains unaffected with variation in the volume fraction of the
inclusions.
The composite grain models with graphite inclusions
were also used by \cite{henning1993} and they did not
find any appreciable shift in the 10$\rm \mu m$ feature with inclusions, 
while \cite{min07} have found that it shifts shortwards.
Our results on the porous silicate grains do not show any shift either in the 10 or the 18$\rm \mu m$
feature. \cite{li} have used porous silicate grain models to study the 10$\rm \mu m$ feature
in the AGNs and found that the 10$\rm \mu m$ peak shifts towards longer wavelength.
For an elaborate comparison of models/results
on silicate IR emission features see e.g. \cite{kessler2006}
and \cite{vaidya2011}. 

(2) We have also found that the composite grain models
with the SiC inclusions modify the absorption profile 
in the spectral region 5--25$\rm \mu m$ and the 10$\rm \mu m$ silicate feature
and do not fit M-type stars. This is consistent with the results obtained by
 \cite{smolder}). 
These results on the composite grain models with SiC inclusions may be also compared 
with the C--rich stars with silicate emission features at 10 and 18$\rm \mu m$
as suggested by \cite{little86} \& \cite{smolder}.
We find that the shape of the SiC feature at 11.3$\rm \mu m$ 
varies with the shape of the grain. This is in agreement with the result 
obtained by \cite{min08}.

We note here that the composite grain models with the inclusions of graphite or SiC
do not seem to fit well the selected sample of Spitzer stars.  
We need to analyze larger sample of M-type and AGB stars observed by Spitzer.

(3) We have found that the composite grain models with number of dipoles N=14440 (AR=2.0) and
graphite inclusions with the volume fractions between 0.1 and 0.3 provide better fit than 
either with SiC inclusions or porous silicate grains for the selected IRAS stars 
(see Tables 1 and 2). \cite{smolder} have used silicates and gehlenite
to fit the observed 10 and 18$\rm \mu m$ silicate features and have also noted that 
grain models with a component of SiC may not fit the observed silicate features to the M-type stars.
The IRAS 17423-1755 is an O-rich AGB star and it too does not fit well with grain models
with SiC inclusion (see more on this AGB star in \cite{mant}).

(4) The composite grain models with graphite and SiC inclusions and porous silicates 
give dust temperatures in the range 250--400K, which fit most of the observed IRAS-LRS curves
selected for this paper. 
This temperature range compares well with the temperature range
suggested by \cite{bowey} and \cite{vosh2008}.
For stars giving good fits with composite grain models with graphite \& SiC inclusions have 
dust temperature range between 270 and 300K, while for those giving good fits with porous models,
have higher dust temperatures between 300 and 370K. The error in temperature estimates
is within 10K.

(5) Flux ratios R is in the range from 0.15 to 0.25 in the dust temperature range of 250--300K,
obtained for the composite grain models with graphite \& SiC inclusions 
and porous silicate grain models compare well
with the observed ratio. These values are lower than that
given by \cite{ossen92}, which are in the range 0.30 to 0.60. The low value of
R may be due to
O-deficient silicates as noted by \cite{little} and 
\cite{ossen}. Most of the stars have the best fit flux ratio R $\sim$ 0.15--0.20 for all
the composite grain models. 

The present study can be extended to Class 12 IRAS objects \cite{gupta2004} which
show similar but weaker silicate emission features at 10$\rm \mu m$. These results on the composite
grain models and earlier studies \cite{little86} and \cite{waters}
clearly indicate that
the composite grain models with other carbonaceous materials (e.g. amorphous silicates 
\& amorphous carbon, ices \& PAHs) as inclusions are required to interpret the observed 
IR emission from the circumstellar
dust around O--rich as well as C--rich stars, to obtain more exact composition, sizes and shapes
of the silicate grains in the circumstellar dust.

We also note that the optical constants for other carbonaceous materials 
by laboratory studies are required.

\section*{Acknowledgments}

D. B. Vaidya thanks IUCAA for funding the visits for completion of this work.
The research has made use of the SIMBAD database, operated at CDS, Strasbourg, France.
\def\aj{AJ}%
\def\actaa{Acta Astron.}%
\def\araa{ARA\&A}%
\def\apj{ApJ}%
\def\apjl{ApJ}%
\def\apjs{ApJS}%
\def\ao{Appl.~Opt.}%
\def\apss{Ap\&SS}%
\def\aap{A\&A}%
\def\aapr{A\&A~Rev.}%
\def\aaps{A\&AS}%
\def\azh{A$Z$h}%
\def\baas{BAAS}%
\def\bac{Bull. astr. Inst. Czechosl.}%
\def\caa{Chinese Astron. Astrophys.}%
\def\cjaa{Chinese J. Astron. Astrophys.}%
\def\icarus{Icarus}%
\def\jcap{J. Cosmology Astropart. Phys.}%
\def\jrasc{JRASC}%
\def\mnras{MNRAS}%
\def\memras{MmRAS}%
\def\na{New A}%
\def\nar{New A Rev.}%
\def\pasa{PASA}%
\def\pra{Phys.~Rev.~A}%
\def\prb{Phys.~Rev.~B}%
\def\prc{Phys.~Rev.~C}%
\def\prd{Phys.~Rev.~D}%
\def\pre{Phys.~Rev.~E}%
\def\prl{Phys.~Rev.~Lett.}%
\def\pasp{PASP}%
\def\pasj{PASJ}%
\def\qjras{QJRAS}%
\def\rmxaa{Rev. Mexicana Astron. Astrofis.}%
\def\skytel{S\&T}%
\def\solphys{Sol.~Phys.}%
\def\sovast{Soviet~Ast.}%
\def\ssr{Space~Sci.~Rev.}%
\def\zap{$Z$Ap}%
\def\nat{Nature}%
\def\iaucirc{IAU~Circ.}%
\def\aplett{Astrophys.~Lett.}%
\def\apspr{Astrophys.~Space~Phys.~Res.}%
\def\bain{Bull.~Astron.~Inst.~Netherlands}%
\def\fcp{Fund.~Cosmic~Phys.}%
\def\gca{Geochim.~Cosmochim.~Acta}%
\def\grl{Geophys.~Res.~Lett.}%
\def\jcp{J.~Chem.~Phys.}%
\def\jgr{J.~Geophys.~Res.}%
\def\jqsrt{J.~Quant.~Spec.~Radiat.~Transf.}%
\def\memsai{Mem.~Soc.~Astron.~Italiana}%
\def\nphysa{Nucl.~Phys.~A}%
\def\physrep{Phys.~Rep.}%
\def\physscr{Phys.~Scr}%
\def\planss{Planet.~Space~Sci.}%
\def\procspie{Proc.~SPIE}%
\let\astap=\aap
\let\apjlett=\apjl
\let\apjsupp=\apjs
\let\applopt=\ao
\bibliographystyle{mn2e}
\bibliography{gupta}

\begin{thebibliography}{}

\bibitem[\protect\citeauthoryear{{Aitken}, {Roche}, {Spenser} \&
  {Jones}}{{Aitken} et~al.}{1979}]{aitken}
{Aitken} D.~K.,  {Roche} P.~F.,  {Spenser} P.~M.,    {Jones} B.,  1979, \apj,
  233, 925

\bibitem[\protect\citeauthoryear{{Backman}, {Witteborn} \& {Gillett}}{{Backman}
  et~al.}{1992}]{back}
{Backman} D.~E.,  {Witteborn} F.~C.,    {Gillett} F.~C.,  1992, \apj, 385, 670

\bibitem[\protect\citeauthoryear{{Bazell} \& {Dwek}}{{Bazell} \&
  {Dwek}}{1990}]{bazel}
{Bazell} D.,  {Dwek} E.,  1990, \apj, 360, 142

\bibitem[\protect\citeauthoryear{{Bode}}{{Bode}}{1988}]{bode88}
{Bode} M.~F.,  1988, in {Bailey} M.~E.,  {Williams} D.~A.,  eds, Dust in the
  Universe {Observations and modelling of circumstellar dust}.
pp 73--102

\bibitem[\protect\citeauthoryear{{Bohren} \& {Huffman}}{{Bohren} \&
  {Huffman}}{1983}]{bohren}
{Bohren} C.~F.,  {Huffman} D.~R.,  1983, {Absorption and scattering of light by
  small particles}.
New York: Wiley, 1983

\bibitem[\protect\citeauthoryear{{Bowey} \& {Adamson}}{{Bowey} \&
  {Adamson}}{2001}]{bowey}
{Bowey} J.~E.,  {Adamson} A.~J.,  2001, \mnras, 320, 131

\bibitem[\protect\citeauthoryear{{Brownlee}}{{Brownlee}}{1987}]{brown}
{Brownlee} D.~E.,  1987, in {Hollenbach} D.~J.,  {Thronson} Jr. H.~A.,  eds,
  Interstellar Processes Vol.~134 of Astrophysics and Space Science Library,
  {Interstellar grains in the solar system}.
pp 513--530

\bibitem[\protect\citeauthoryear{{Carciofi}, {Bjorkman} \&
  {Magalh{\~a}es}}{{Carciofi} et~al.}{2004}]{car}
{Carciofi} A.~C.,  {Bjorkman} J.~E.,    {Magalh{\~a}es} A.~M.,  2004, \apj,
  604, 238

\bibitem[\protect\citeauthoryear{{Ch{\'y}lek}, {Videen}, {Geldart}, {Dobbie} \&
  {Tso}}{{Ch{\'y}lek} et~al.}{2000}]{chylek2000}
{Ch{\'y}lek} P.,  {Videen} G.,  {Geldart} D.~J.~W.,  {Dobbie} J.~S.,    {Tso}
  H.~C.~W.,  2000, {Effective Medium Approximations for Heterogeneous
  Particles}.
Academic Press, San Diego, USA

\bibitem[\protect\citeauthoryear{{Cohen}}{{Cohen}}{1984}]{cohen}
{Cohen} M.,  1984, \mnras, 206, 137

\bibitem[\protect\citeauthoryear{{Draine}}{{Draine}}{1988}]{draine1988}
{Draine} B.~T.,  1988, \apj, 333, 848

\bibitem[\protect\citeauthoryear{{Draine}}{{Draine}}{2003}]{draine2003}
{Draine} B.~T.,  2003, \apj, 598, 1017

\bibitem[\protect\citeauthoryear{{Draine} \& {Flatau}}{{Draine} \&
  {Flatau}}{2003}]{draineflatau2003}
{Draine} B.~T.,  {Flatau} P.~J.,  2003, ArXiv Astrophysics e-prints

\bibitem[\protect\citeauthoryear{{Draine} \& {Lee}}{{Draine} \&
  {Lee}}{1984}]{draine1984}
{Draine} B.~T.,  {Lee} H.~M.,  1984, \apj, 285, 89

\bibitem[\protect\citeauthoryear{{Draine} \& {Li}}{{Draine} \&
  {Li}}{2007}]{draine-li}
{Draine} B.~T.,  {Li} A.,  2007, \apj, 657, 810

\bibitem[\protect\citeauthoryear{{Greenberg}}{{Greenberg}}{1990}]{green}
{Greenberg} J.~M.,  1990, {The evidence that comets are made of interstellar
  dust}.
Kluwer Academic Publishers, Dordrecht, The Netherlands, pp 99--120

\bibitem[\protect\citeauthoryear{{Groenewegen}}{{Groenewegen}}{1997}]{groen}
{Groenewegen} M.~A.~T.,  1997, \aap, 317, 503

\bibitem[\protect\citeauthoryear{{Gupta}, {Mukai}, {Vaidya}, {Sen} \&
  {Okada}}{{Gupta} et~al.}{2005}]{gupta2005}
{Gupta} R.,  {Mukai} T.,  {Vaidya} D.~B.,  {Sen} A.~K.,    {Okada} Y.,  2005,
  \aap, 441, 555

\bibitem[\protect\citeauthoryear{{Gupta}, {Singh}, {Volk} \& {Kwok}}{{Gupta}
  et~al.}{2004}]{gupta2004}
{Gupta} R.,  {Singh} H.~P.,  {Volk} K.,    {Kwok} S.,  2004, VizieR Online Data
  Catalog, 215

\bibitem[\protect\citeauthoryear{{Henning} \& {Stognienko}}{{Henning} \&
  {Stognienko}}{1993}]{henning1993}
{Henning} T.,  {Stognienko} R.,  1993, \aap, 280, 609

\bibitem[\protect\citeauthoryear{{Houck}, {Roellig}, {van Cleve}, {Forrest},
  {Herter}, {Lawrence}, {Matthews}, {Reitsema}, {Soifer} \& {and
  others}}{{Houck} et~al.}{2004}]{houck}
{Houck} J.~R.,  {Roellig} T.~L.,  {van Cleve} J.,  {Forrest} W.~J.,  {Herter}
  T.,  {Lawrence} C.~R.,  {Matthews} K.,  {Reitsema} H.~J.,  {Soifer} B.~T.,
  {and others} 2004, \apjs, 154, 18

\bibitem[\protect\citeauthoryear{{Iat{\`i}}, {Giusto}, {Saija}, {Borghese},
  {Denti}, {Cecchi-Pestellini} \& {Aiello}}{{Iat{\`i}} et~al.}{2004}]{iati}
{Iat{\`i}} M.~A.,  {Giusto} A.,  {Saija} R.,  {Borghese} F.,  {Denti} P.,
  {Cecchi-Pestellini} C.,    {Aiello} S.,  2004, \apj, 615, 286

\bibitem[\protect\citeauthoryear{{Kessler-Silacci}, {Augereau}, {Dullemond},
  {Geers}, {Lahuis}, {Evans} II, {van Dishoeck}, {Blake}, {Boogert}, {Brown},
  {J{\o}rgensen}, {Knez} \& {Pontoppidan}}{{Kessler-Silacci}
  et~al.}{2006}]{kessler2006}
{Kessler-Silacci} J.,  {Augereau} J.-C.,  {Dullemond} C.~P.,  {Geers} V.,
  {Lahuis} F.,  {Evans} II N.~J.,  {van Dishoeck} E.~F.,  {Blake} G.~A.,
  {Boogert} A.~C.~A.,  {Brown} J.,  {J{\o}rgensen} J.~K.,  {Knez} C.,
  {Pontoppidan} K.~M.,  2006, \apj, 639, 275

\bibitem[\protect\citeauthoryear{{Kim} \& {Martin}}{{Kim} \&
  {Martin}}{1995}]{kim}
{Kim} S.-H.,  {Martin} P.~G.,  1995, \apj, 444, 293

\bibitem[\protect\citeauthoryear{{Kirchschlager} \& {Wolf}}{{Kirchschlager} \&
  {Wolf}}{2014}]{kirch}
{Kirchschlager} F.,  {Wolf} S.,  2014, \aap, 568, A103

\bibitem[\protect\citeauthoryear{{K{\"o}hler} \& {Mann}}{{K{\"o}hler} \&
  {Mann}}{2004}]{kohler}
{K{\"o}hler} M.,  {Mann} I.,  2004, \jqsrt, 89, 453

\bibitem[\protect\citeauthoryear{{Krivov}, {Mann} \& {Krivova}}{{Krivov}
  et~al.}{2000}]{kriv}
{Krivov} A.~V.,  {Mann} I.,    {Krivova} N.~A.,  2000, \aap, 362, 1127

\bibitem[\protect\citeauthoryear{{Lasue}, {Botet}, {Levasseur-Regourd} \&
  {Hadamcik}}{{Lasue} et~al.}{2009}]{lasue}
{Lasue} J.,  {Botet} R.,  {Levasseur-Regourd} A.~C.,    {Hadamcik} E.,  2009,
  \icarus, 203, 599

\bibitem[\protect\citeauthoryear{{Lebouteiller}, {Barry}, {Goes}, {Sloan},
  {Spoon}, {Weedman}, {Bernard-Salas} \& {Houck}}{{Lebouteiller}
  et~al.}{2015}]{lebout}
{Lebouteiller} V.,  {Barry} D.~J.,  {Goes} C.,  {Sloan} G.~C.,  {Spoon}
  H.~W.~W.,  {Weedman} D.~W.,  {Bernard-Salas} J.,    {Houck} J.~R.,  2015,
  \apjs, 218, 21

\bibitem[\protect\citeauthoryear{{Levasseur-Regourd} \&
  {Hadamcik}}{{Levasseur-Regourd} \& {Hadamcik}}{2013}]{levas}
{Levasseur-Regourd} A.,  {Hadamcik} E.,  2013, in AAS/Division for Planetary
  Sciences Meeting Abstracts Vol.~45 of AAS/Division for Planetary Sciences
  Meeting Abstracts, {Properties of dust in inner cometary comae from
  polarimetric observations}.
p. 505.02

\bibitem[\protect\citeauthoryear{{Li} \& {Greenberg}}{{Li} \&
  {Greenberg}}{1998}]{li1998}
{Li} A.,  {Greenberg} J.~M.,  1998, \aap, 331, 291

\bibitem[\protect\citeauthoryear{{Li}, {Shi} \& {Li}}{{Li} et~al.}{2008}]{li}
{Li} M.~P.,  {Shi} Q.~J.,    {Li} A.,  2008, \mnras, 391, L49

\bibitem[\protect\citeauthoryear{{Little-Marenin}}{{Little-Marenin}}{1986}]{little86}
{Little-Marenin} I.~R.,  1986, \apjl, 307, L15

\bibitem[\protect\citeauthoryear{{Little-Marenin} \& {Little}}{{Little-Marenin}
  \& {Little}}{1990}]{little}
{Little-Marenin} I.~R.,  {Little} S.~J.,  1990, \aj, 99, 1173

\bibitem[\protect\citeauthoryear{{Lorenz-Martins} \&
  {Lefevre}}{{Lorenz-Martins} \& {Lefevre}}{1993}]{lorenz}
{Lorenz-Martins} S.,  {Lefevre} J.,  1993, \aap, 280, 567

\bibitem[\protect\citeauthoryear{{Manteiga}, {Garc{\'{\i}}a-Hern{\'a}ndez},
  {Ulla}, {Manchado} \& {Garc{\'{\i}}a-Lario}}{{Manteiga} et~al.}{2011}]{mant}
{Manteiga} M.,  {Garc{\'{\i}}a-Hern{\'a}ndez} D.~A.,  {Ulla} A.,  {Manchado}
  A.,    {Garc{\'{\i}}a-Lario} P.,  2011, \aj, 141, 80

\bibitem[\protect\citeauthoryear{{Mathis}}{{Mathis}}{1996}]{mathis}
{Mathis} J.~S.,  1996, \apj, 472, 643

\bibitem[\protect\citeauthoryear{{Mathis}, {Rumpl} \& {Nordsieck}}{{Mathis}
  et~al.}{1977}]{mathis1977}
{Mathis} J.~S.,  {Rumpl} W.,    {Nordsieck} K.~H.,  1977, \apj, 217, 425

\bibitem[\protect\citeauthoryear{{Min}, {Hovenier}, {de Koter}, {Waters} \&
  {Dominik}}{{Min} et~al.}{2005}]{min05}
{Min} M.,  {Hovenier} J.~W.,  {de Koter} A.,  {Waters} L.~B.~F.~M.,
  {Dominik} C.,  2005, \icarus, 179, 158

\bibitem[\protect\citeauthoryear{{Min}, {Hovenier}, {Waters} \& {de
  Koter}}{{Min} et~al.}{2008}]{min08}
{Min} M.,  {Hovenier} J.~W.,  {Waters} L.~B.~F.~M.,    {de Koter} A.,  2008,
  \aap, 489, 135

\bibitem[\protect\citeauthoryear{{Min}, {Waters}, {de Koter}, {Hovenier},
  {Keller} \& {Markwick-Kemper}}{{Min} et~al.}{2007}]{min07}
{Min} M.,  {Waters} L.~B.~F.~M.,  {de Koter} A.,  {Hovenier} J.~W.,  {Keller}
  L.~P.,    {Markwick-Kemper} F.,  2007, \aap, 462, 667

\bibitem[\protect\citeauthoryear{{Mishchenko}, {Travis} \&
  {Lacis}}{{Mishchenko} et~al.}{2002}]{mish2002}
{Mishchenko} M.~L.,  {Travis} L.~D.,    {Lacis} A.~A.,  2002, {Scattering,
  absorption, and emission of light by small particles}.
Cambridge University Press, UK

\bibitem[\protect\citeauthoryear{{O'Donnell}}{{O'Donnell}}{1994}]{odon}
{O'Donnell} J.~E.,  1994, \apj, 437, 262

\bibitem[\protect\citeauthoryear{{Olnon}, {Raimond}, {Neugebauer}, {van
  Duinen}, {Habing}, {Aumann}, {Beintema} \& {and others}}{{Olnon}
  et~al.}{1986}]{olnon}
{Olnon} F.~M.,  {Raimond} E.,  {Neugebauer} G.,  {van Duinen} R.~J.,  {Habing}
  H.~J.,  {Aumann} H.~H.,  {Beintema} D.~A.,    {and others} 1986, \aaps, 65,
  607

\bibitem[\protect\citeauthoryear{{Ossenkopf}}{{Ossenkopf}}{1991}]{ossen}
{Ossenkopf} V.,  1991, \aap, 251, 210

\bibitem[\protect\citeauthoryear{{Ossenkopf}, {Henning} \&
  {Mathis}}{{Ossenkopf} et~al.}{1992}]{ossen92}
{Ossenkopf} V.,  {Henning} T.,    {Mathis} J.~S.,  1992, \aap, 261, 567

\bibitem[\protect\citeauthoryear{{Pegourie}}{{Pegourie}}{1988}]{pegourie1988}
{Pegourie} B.,  1988, \aap, 194, 335

\bibitem[\protect\citeauthoryear{{Perrin} \& {Lamy}}{{Perrin} \&
  {Lamy}}{1990}]{perrin90a}
{Perrin} J.-M.,  {Lamy} P.~L.,  1990, \apj, 364, 146

\bibitem[\protect\citeauthoryear{{Perrin} \& {Sivan}}{{Perrin} \&
  {Sivan}}{1990}]{perrin90b}
{Perrin} J.-M.,  {Sivan} J.-P.,  1990, \aap, 228, 238

\bibitem[\protect\citeauthoryear{{Purcell} \& {Pennypacker}}{{Purcell} \&
  {Pennypacker}}{1973}]{purcell}
{Purcell} E.~M.,  {Pennypacker} C.~R.,  1973, \apj, 186, 705

\bibitem[\protect\citeauthoryear{{Saija}, {Iat{\`i}}, {Borghese}, {Denti},
  {Aiello} \& {Cecchi-Pestellini}}{{Saija} et~al.}{2001}]{saija2001}
{Saija} R.,  {Iat{\`i}} M.~A.,  {Borghese} F.,  {Denti} P.,  {Aiello} S.,
  {Cecchi-Pestellini} C.,  2001, \apj, 559, 993

\bibitem[\protect\citeauthoryear{{Siebenmorgen}, {Voshchinnikov} \&
  {Bagnulo}}{{Siebenmorgen} et~al.}{2014}]{siber}
{Siebenmorgen} R.,  {Voshchinnikov} N.~V.,    {Bagnulo} S.,  2014, \aap, 561,
  A82

\bibitem[\protect\citeauthoryear{{Simpson}}{{Simpson}}{1991}]{simp}
{Simpson} J.~P.,  1991, \apj, 368, 570

\bibitem[\protect\citeauthoryear{{Smolders}, {Neyskens}, {Blommaert}, {Hony},
  {van Winckel}, {Decin}, {van Eck}, {Sloan}, {Cami}, {Uttenthaler}, {Degroote}
  \& {and others}}{{Smolders} et~al.}{2012}]{smolder}
{Smolders} K.,  {Neyskens} P.,  {Blommaert} J.~A.~D.~L.,  {Hony} S.,  {van
  Winckel} H.,  {Decin} L.,  {van Eck} S.,  {Sloan} G.~C.,  {Cami} J.,
  {Uttenthaler} S.,  {Degroote} P.,    {and others} 2012, \aap, 540, A72

\bibitem[\protect\citeauthoryear{{Sylvester}, {Skinner}, {Barlow} \&
  {Mannings}}{{Sylvester} et~al.}{1996}]{sylv}
{Sylvester} R.~J.,  {Skinner} C.~J.,  {Barlow} M.~J.,    {Mannings} V.,  1996,
  \mnras, 279, 915

\bibitem[\protect\citeauthoryear{{Telesco} \& {Knacke}}{{Telesco} \&
  {Knacke}}{1991}]{teles}
{Telesco} C.~M.,  {Knacke} R.~F.,  1991, \apjl, 372, L29

\bibitem[\protect\citeauthoryear{{Vaidya} \& {Gupta}}{{Vaidya} \&
  {Gupta}}{1997}]{vaidya1997}
{Vaidya} D.~B.,  {Gupta} R.,  1997, \aap, 328, 634

\bibitem[\protect\citeauthoryear{{Vaidya} \& {Gupta}}{{Vaidya} \&
  {Gupta}}{2009}]{vaidya2009}
{Vaidya} D.~B.,  {Gupta} R.,  2009, \jqsrt, 110, 1726

\bibitem[\protect\citeauthoryear{{Vaidya} \& {Gupta}}{{Vaidya} \&
  {Gupta}}{2011}]{vaidya2011}
{Vaidya} D.~B.,  {Gupta} R.,  2011, \aap, 528, A57

\bibitem[\protect\citeauthoryear{{Vaidya}, {Gupta}, {Dobbie} \&
  {Chylek}}{{Vaidya} et~al.}{2001}]{vaidya2001}
{Vaidya} D.~B.,  {Gupta} R.,  {Dobbie} J.~S.,    {Chylek} P.,  2001, \aap, 375,
  584

\bibitem[\protect\citeauthoryear{{Vaidya}, {Gupta} \& {Snow}}{{Vaidya}
  et~al.}{2007}]{vaidya2007}
{Vaidya} D.~B.,  {Gupta} R.,    {Snow} T.~P.,  2007, \mnras, 379, 791

\bibitem[\protect\citeauthoryear{{Voshchinnikov}, {Das}, {Yakovlev} \&
  {Il'in}}{{Voshchinnikov} et~al.}{2013}]{vosh2013}
{Voshchinnikov} N.~V.,  {Das} H.~K.,  {Yakovlev} I.~S.,    {Il'in} V.~B.,
  2013, Astronomy Letters, 39, 421

\bibitem[\protect\citeauthoryear{{Voshchinnikov} \& {Henning}}{{Voshchinnikov}
  \& {Henning}}{2008}]{vosh2008}
{Voshchinnikov} N.~V.,  {Henning} T.,  2008, \aap, 483, L9

\bibitem[\protect\citeauthoryear{{Voshchinnikov}, {Il'in} \&
  {Henning}}{{Voshchinnikov} et~al.}{2005}]{vosh2005}
{Voshchinnikov} N.~V.,  {Il'in} V.~B.,    {Henning} T.,  2005, \aap, 429, 371

\bibitem[\protect\citeauthoryear{{Voshchinnikov}, {Il'in}, {Henning} \&
  {Dubkova}}{{Voshchinnikov} et~al.}{2006}]{vosh2006}
{Voshchinnikov} N.~V.,  {Il'in} V.~B.,  {Henning} T.,    {Dubkova} D.~N.,
  2006, \aap, 445, 167

\bibitem[\protect\citeauthoryear{{Waters}, {Beintema}, {Zijlstra}, {de Koter},
  {Molster}, {Bouwman}, {de Jong}, {Pottasch} \& {de Graauw}}{{Waters}
  et~al.}{1998}]{waters}
{Waters} L.~B.~F.~M.,  {Beintema} D.~A.,  {Zijlstra} A.~A.,  {de Koter} A.,
  {Molster} F.~J.,  {Bouwman} J.,  {de Jong} T.,  {Pottasch} S.~R.,    {de
  Graauw} T.,  1998, \aap, 331, L61

\bibitem[\protect\citeauthoryear{{Werner}, {Roellig}, {Low}, {Rieke}, {Rieke},
  {Hoffmann}, {Young}, {Houck} \& {and others}}{{Werner} et~al.}{2004}]{werner}
{Werner} M.~W.,  {Roellig} T.~L.,  {Low} F.~J.,  {Rieke} G.~H.,  {Rieke} M.,
  {Hoffmann} W.~F.,  {Young} E.,  {Houck} J.~R.,    {and others} 2004, \apjs,
  154, 1

\bibitem[\protect\citeauthoryear{{Whittet}}{{Whittet}}{2003}]{whitt}
{Whittet} D.~C.~B.,  2003, {Dust in the galactic environment}.
Institute of Physics (IOP) Publishing, 2003 Series in Astronomy and
  Astrophysics, ISBN 0750306246, Bristol

\bibitem[\protect\citeauthoryear{{Wolff}, {Clayton} \& {Gibson}}{{Wolff}
  et~al.}{1998}]{wolff1998}
{Wolff} M.~J.,  {Clayton} G.~C.,    {Gibson} S.~J.,  1998, \apj, 503, 815

\bibitem[\protect\citeauthoryear{{Wolff}, {Clayton}, {Martin} \&
  {Schulte-Ladbeck}}{{Wolff} et~al.}{1994}]{wolff1994}
{Wolff} M.~J.,  {Clayton} G.~C.,  {Martin} P.~G.,    {Schulte-Ladbeck} R.~E.,
  1994, \apj, 423, 412

\bibitem[\protect\citeauthoryear{{Woolf}}{{Woolf}}{1973}]{woolf73}
{Woolf} N.~J.,  1973, in {Greenberg} J.~M.,  {van de Hulst} H.~C.,  eds,
  Interstellar Dust and Related Topics Vol.~52 of IAU Symposium, {Circumstellar
  Infrared Emission. I the Circumstellar Origin of Interstellar Dust (review)}.
p.~485

\bibitem[\protect\citeauthoryear{{Woolf} \& {Ney}}{{Woolf} \&
  {Ney}}{1969}]{woolf69}
{Woolf} N.~J.,  {Ney} E.~P.,  1969, \apjl, 155, L181

\bibitem[\protect\citeauthoryear{{Zubko}, {Dwek} \& {Arendt}}{{Zubko}
  et~al.}{2004}]{zubko}
{Zubko} V.,  {Dwek} E.,    {Arendt} R.~G.,  2004, \apjs, 152, 211

\end{thebibliography}

\end{document}